# NON-CAUSAL FIR FILTERS FOR THE MAXIMUM RETURN FROM CAPITAL MARKETS

ANDRZEJ DYKA ♣

Gdańsk University of Technology
Narutowicza 11/12, 80-952 Gdańsk, Poland

In this paper we consider a trading strategy, which consists in buying or selling a financial instrument when the smoothing, non-causal FIR (Final Impulse Response) filter output attains a local minimum or maximum, respectively. Upon tis assumption the goal of this paper is to determine the "best" non-causal smoothing FIR filters, which provide maximum value of the return from the market. The assumed non-causality is obtained by advancing the output signal to compensate for the delay introduced by the *a priori* known filter. The best result were obtained for the impulse response given by the Pascal triangle and the family of symmetric power triangles, both for the case of trading with, and without the transaction fee. It was found that the transaction fee dramatically reduces a possible net return from the market, and therefore should not be omitted in market analyzes.

## 1. Objective

Low-pass FIR filtering commonly used in the analysis of markets is the cause of inevitable changes in the original data. The desirable change, which is a rationale of such filtering, consists in smoothing input data. It is believed that this positive change enables for a better estimation of market movements. On the other hand, the undesirable change in the original data consists in delaying the filter output to such an extent that anticipations of future market movement are hardly possible. A tradeoff between the degree of desirable smoothing and undesirable delay provides a very little room for optimizing such filters, e.g., Dyka & Kaźmierczak, [1]. The main goal of this contribution is to determine the "best" non-causal smoothing FIR filters, which provide maximum of the net return from the market. The assumed non-causality is obtained by advancing the output signal to compensate for the delay introduced by the *a priori* known filter. This way the smoothed data is not delayed with respect to the original data. Upon this assumption the following trading criterion is assumed: the position is opened ( i.e. the financial instrument is bought) when the derivative of the filter output signal upwardly crosses zero. The position is closed ( i.e. the the financial instrument is sold) when the derivative of the filter output signal downwardly crosses zero. The objective of this contribution is to determine the "best" filters which yield maximum return for real market scenarios.

## 2. Fundamentals of linear FIR filters

In general linear systems can be described by linear differential equations with constant coefficients *e.g.,* [2]. The term linearity denotes two important properties i.e., superposition, and frequency preservation principle. The principle of superposition says that the response of a linear filter due to a number of inputs applied simultaneously is equal to the sum of its responses to each input applied separately. The principle of frequency preservation says that the response or output signal of the filter contains only these frequencies which are represented in the input signal.

---

♣ www.dyka.info.pl

Any linear filter is uniquely described by a specific time domain function referred to as the impulse response h(t). Impulse response h(t) is by the definition the response of the filter to the Dirac's delta function applied to input.

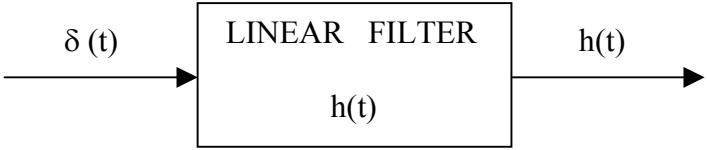

Fig. 1. Definition of filter impulse response h(t)

Dirac's delta is is an idealized function, not feasible in any real world. In practice a number of reasonable approximations and variants of the above scheme are possible. For this reason the above scheme represents a powerful tool for a quick assesment of impulse response of linear systems used in numerous applications of sience and technology.

In a general case the filtering equation, which describes the relationship between output and input signal takes the following form

$$y(t) = x(t) * h(t), \qquad (1)$$

where the asterisk $*$ denotes convolution.

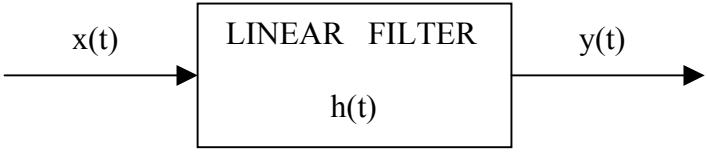

Fig. 2. Representation of linear filter – a general case

Consequently, convolution is defined, as follows,

$$y(t) = \int_{-\infty}^{\infty} x(\tau) \cdot h(t - \tau) \, d\tau, \qquad (2)$$

and defines linear filtering in the most general way.

A sub-class of impulse responses, limited in argument, in the discrete time domain are termed FIR – Finite Impulse Response, whereas filtering performed by FIR filters in the discrete time domain is usually referred to as the 'digital filtering'. In the discrete time domain the integral in (2) takes the form of an appropriate sum, and the infinite limits of integration are replaced by finite numbers resulting from the length of both, discrete input sequence x and the discrete impulse response h. Finally, it is worth to emphasize that for any input signal x of length T and impulse response h of length q the output signal length equals T+q-1. If T > q then first and last q samples in the output signal are referred to as the 'transition state' whereas remaining T-q+1 samples are termed the 'steady state'. For T=q the steady state is represented by a single sample, for T<q - does not exist.

## 3. Assumptions and definition of variables

It is assumed here, that the output signal of a low-pass FIR filter is continuously differentiable, which in practice is always met. Hence, one may conclude that maxima and minima in the output signal of a low-pass FIR filter appear alternately. Therefore, each pair of the two consecutive extrema i.e maximum preceded by minimum will be referred to as the completed transaction.

In addition the following assumptions are to be made:

1. The amounts of instrument bought and sold for a completed transaction are equal.

2. The stock liquidity is much higher than the transaction level, which means that ask and bid volume is much higher than the amount of assumed transaction.

In other words, both above assumptions mean that at every transaction we buy and sell a single unit of the instrument, and that these transactions do not affect the price.

For computations the database of one–minute quotations for the futures on WIG 20 index, covering period of time from October 30, 2001 thru June 16, 2003, that is about 145000 samples was used. During this period of time the average transaction fee offered by brokers was between 10 –14 PLN. Preliminary computations proved that for varying length of impulse response only symmetric shapes of impulse response (even function) could provide a positive net return for a wide range of the length q of filter impulse response. Therefore, for further analysis we assumed filters for which impulse response is an even function and have odd number of discrete samples. Consequently, we assumed a discrete variable n varying from n=1 to n=199. This corresponds to filter length q given by even numbers from q=3 to q= 399, as follows:

$$q = 2*n+1 \qquad (3)$$

For every value of n=1:199 the return function R(n) was computed. R(n) is the sequence, which has r(n) of non-zero numbers, each representing an outcome (profit or loss) of the completed transaction. Assuming one-minute quotations of k data samples we define the following:

$$\mathbf{N}(n) = \sum_{i=1}^{r(n)} R(n) / k \qquad \text{- net return per minute} \qquad (4)$$

$$\mathbf{L}(n) = \sum_{i=1}^{r(n)} [ R(n)* (R(n)<0) ] / k \qquad \text{- loss per minute} \qquad (5)$$

$$\mathbf{F}(n) = [ F * r(n) ] / k \qquad \text{- fee per minute,} \qquad (6)$$

where F is the fee for a completed transaction.

By the virtue of (4) and (5) we have:

$$G(n) = N(n) - L(n) \quad \text{- gross return per minute} \quad (7)$$

In addition define the following:

1. Filter efficiency without fee,

$$E(n) = 100*(N(n)/G(n)), \quad [\%] \quad (8)$$

2. Filter sensitivity to fee,

$$S(n) = 100*(F(n)/G(n)), \quad [\%] \quad (9)$$

### 3. Computations and results

The computations were carried out for the idealized, academic case without including the transaction fee, and the real case including the 12 PLN real transaction fee. Several finite length even functions, representing impulse response of the FIR filter were taken into consideration, including the most well-known windows used in the area of spectrum analysis and antenna theory. It was found that the best results were obtained for the two following cases of the form of impulse response:

1. The superior performance in terms of maximum of average net return was obtained for the impulse response given by Pascal triangle i.e. for

$$h(t) = \binom{2n}{n+t} \quad (10)$$

2. A very good performance, was also obtained for the family of symmetric power triangles for which the left side is a function

$$h(t) = a*t \char`\^ p \quad (11)$$

where a is constant.

For the family of symmetric power triangles it was found that the most interesting performance can be obtained for the values of p varying from 1 to 3. For p = 1 we have isosceles triangle, which has previously been recommended by Dyka & Kaźmierczak for filtering market data, [1]. In the the case of p ≈1.38 the filter was found to generate the minimum of average loss against average return. It was also found that the commonly used Moving Average (MA), which is equivalent to the rectangular shape of impulse response offers a relatively poor performance. The inferior performance of MA in comparison to other impulse responses has already been indicated by Dyka & Kaźmierczak, [1].

The graphs of the net return for the impulse response given by isosceles triangle of p=1, power triangle of p=.38, Pascal triangle, and rectangle (MA) have been presented in Fig 1, and Fig 2.

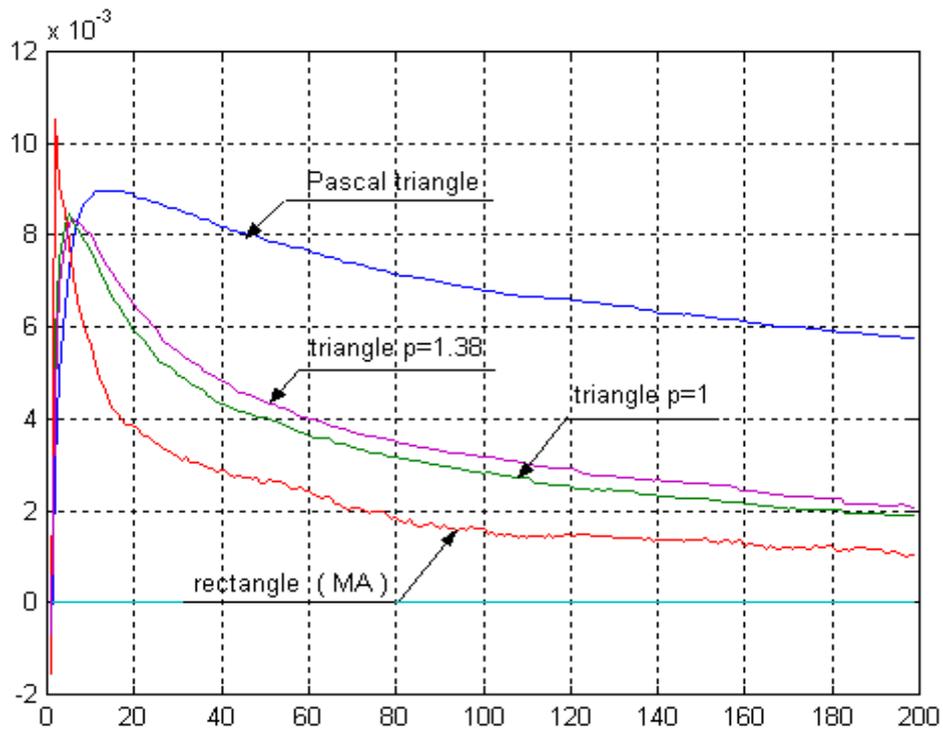

Fig 3. **N**(n) - net return without transaction fee versus n

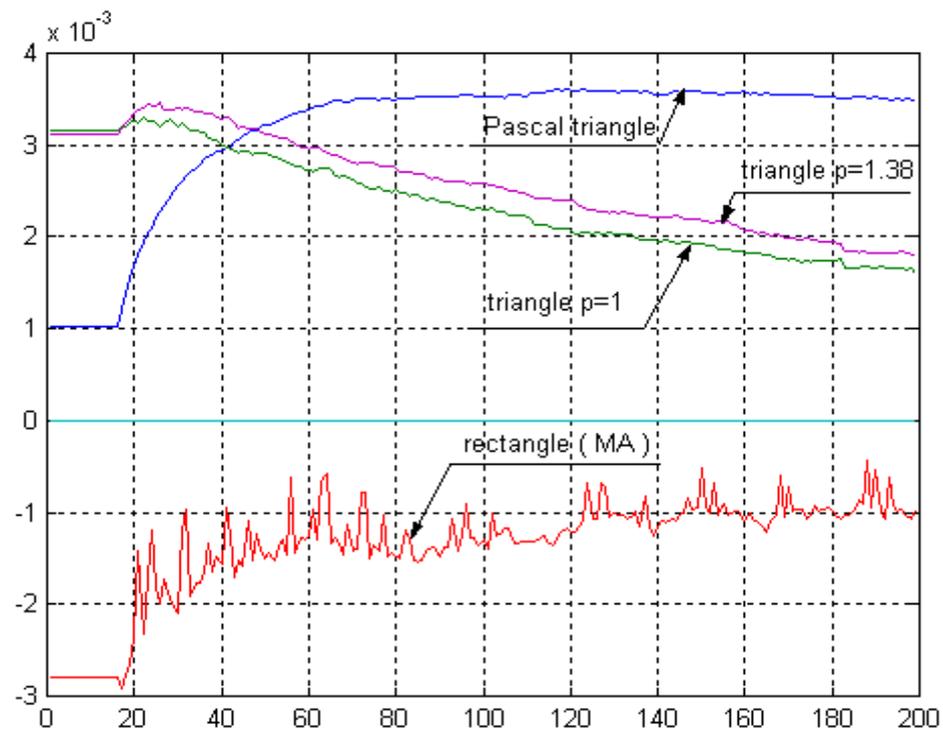

Fig 4. **N**(n) – **F**(n) - net return including a real transaction fee versus n

As can be seen in Fig 3 all cases of the analyzed impulse response shape show a positive net return for almost all values of n. Pascal triangle, except for n < 7, provides a superior performance compared to other impulse response shapes, irrespctive of n. When the real transaction fee is included into analysis ( Fig. 4). the situation dramatically changes. Needless to say, a significant drop in net return for all cases of impulse response is observed. Pascal triangle is still superior over other impulse responses for n > 40, and n > 48, in the case of isosceles triangle ( p=1 ) and the power triangle ( p=1.38 ), respectively. For lesser values of n both of impulse responses mentioned above yield a return higher than Pascal triangle. This is due to the fact that Pascal triangle generates much higher number of transactions than other filters. Therefore, for lower values of n when the number of transactions is very high, the negative impact of the transaction fee on the Pascal triangle performance is much stronger, than on the other two. It is worth to emphasize that the rectangular shape of impulse response (Moving Average) completely fails providing positive net return in the presence of transaction fee, generating losses for any value of n.

For a deeper insight into results the net return $N(n)$ ,(4), filter efficiency without fee $E(n)$, (8), and filter sensitivity to fee $S(n)$, (9), were averaged for n =15 to n=199. The resulting norms have synthetically been presented in Table 1.

Table 1

|  | Without transaction fee | | Including transaction fee | |
|---|---|---|---|---|
|  | Average $N(n)$ [ % ] / minute | Average $E(n)$ [ % ] | Average $N(n)$- $F(n)$ [ % ] / minute | Average $S(n)$ [ % ] |
| p=1 | 0.3072 | 99.62 | 0.2320 | **24.38** |
| p=1.38 | 0.3420 | **99.75** | 0.2548 | 25.42 |
| Pascal triangle | **0.6957** | 99.44 | **0.3318** | 52.01 |

As shown in Table 1 the Pascal triangle impulse response yields maximum and positive average net return either without or with the real transaction fee equal to 0.6957 [%] / min and 0.3318 [%] / min, respectively. The maximum average efficiency of 99.75 [%] is obtained for the triangle case of p = 1.38. Admittedly, other filters under consideration only slightly detract from this value. The minimum of average filter sensitivity to fee of 24.38 [%] is attained in the case of p = 1, i.e. for the isosceles triangle.

## 5. Concluding remarks

It was found, that in the idealized case i.e. without the transaction fee, practically all filters provide positive return from the market. However, when the real transaction fee is included, only some of them provide positive return, whereas other generate loss. The following conclusions can be drawn from the above presented survey.

1. The maximum of the average net return has been obtained for the impulse response described by Pascal triangle. However one cannot rule out that other shapes of impulse response may locally provide a better result.

2. The quantitative results for best filters, though non-causal, and therefore not feasible can be considered as a reference or benchmark for other comparative analyses, such as for instance curve fitting, etc. At least they determines the maximum return from the market, which probably will never be attained in practice.

3. The real transaction fee has a very strong impact on the final net return from the market. This is an important remark because sometimes analysts or researchers "for the sake of clarity" or "simplicity" tend to assume the transaction fee to be zero.

4. Moving average (MA) i.e., rectangular impulse response (boxcar) commonly used by market analysts dramatically fails providing positive net return when the real transaction fee is taken into account.

<div style="text-align:center">REFERENCES</div>